\documentstyle[12pt]{article}

\newcommand{\be}{\begin{eqnarray}}
\newcommand{\ee}{\end{eqnarray}}
\newcommand{\balpha}{\mbox{\boldmath $\alpha$}}
\newcommand{\bgamma}{\mbox{\boldmath $\gamma$}}
\newcommand{\bsigma}{\mbox{\boldmath $\sigma$}}
\begin{document}

\title{\bf Quasi-exact Solvability of  Dirac
Equations\footnote{Based on talks presented at the 11th
International Conference on Symmetry Methods in Physics (Jun
21-24, 2004, Prague) and at the XXIII International Conference on
Differential Geometric Methods in Theoretical Physics (Aug 20-26,
2005, Nankai, Tianjin, China).}}

\author{Choon-Lin Ho
\\ {\small{\sl Department of Physics, Tamkang University, Tamsui 25137,
Taiwan, R.O.C.}} }

\date{}

\maketitle

\vskip 1cm

\begin{abstract}
We present a general procedure for determining quasi-exact
solvability of the Dirac and the Pauli  equation with an
underlying $sl(2)$ symmetry. This procedure makes full use of the
close connection between quasi-exactly solvable systems and
supersymmetry.  The Dirac-Pauli equation with spherical electric
field is taken as an example to illustrate the procedure.
\end{abstract}

\vskip 1truecm

{\bf 1.} In this talk we present a general procedure for
determining quasi-exact solvability of the Dirac and the Pauli
equation with an underlying $sl(2)$ symmetry. This procedure makes
full use of the close connection between quasi-exactly solvable
(QES) systems and supersymmetry (SUSY), or equivalently, the
factorizability of the equation. Based on this procedure, we have
demonstrated that the Pauli and the Dirac equation coupled
minimally with a vector potential \cite{HoRoy1,BK}, neutral Dirac
particles in external electric fields (which are equivalent to
generalized Dirac oscillators) \cite{HoRoy2,BN}, and Dirac
equation with a Lorentz scalar potential \cite{Ho} are physical
examples of QES systems.

Here we only give the main ideas of the procedures, and refer the
readers to \cite{HoRoy1,HoRoy2,Ho} for details.

\vskip 0.7cm

 {\bf 2.} For all the cases cited above, one can reduce
the corresponding multi-component equations to a set of
one-variable equations possessing one-dimensional SUSY after
separating the variables in a suitable coordinate system.
Typically the set of equations takes the form \be
\left(\frac{d}{dr} + W(r)\right)f_- &=& {\cal E}^+
f_+~,\label{f_-}\\ \left(-\frac{d}{dr} + W(r)\right)f_+ &=& {\cal
E}^ - f_-~, \label{f_+} \ee where $r$ is the basic variable, e.g.
the radial coordinate, and $f_\pm$ are, say, the two components of
the radial part of the Dirac wave function. The superpotential $W$
is related to the external field configuration, and ${\cal E}^\pm$
involve the energy and mass of the particle. We can rewrite this
set of equations as \be A^-A^+f_-&=&\epsilon f_-~,\\ A^+A^-f_+&=&
\epsilon f_+~, \ee with
\be
A^\pm\equiv \pm \frac{d}{dr}  +W~, ~~~\epsilon\equiv {\cal E}^+
{\cal E}^-~. \label{A} \ee Explicitly, the above equations read
\be
\left(-\frac{d^2}{dr^2} + W^2 \mp W^\prime\right)f_\mp = \epsilon
f_\mp~. \label{susy} \ee
 Here and below the prime means
differentiation with respect to the basic variable.
Eq.(\ref{susy}) clearly exhibits the SUSY structure of the system.
The operators acting on $f_\pm$ in Eq.(\ref{susy}) are said to be
factorizable, i.e. as products of $A^-$ and $A^+$. The ground
state, with $\epsilon = 0$,  is given by one of the following two
sets of equations: \be A^+ f_-^{(0)}(r)&=& 0~~,~~~
f_+^{(0)}(r)=0~;\\ A^- f_+^{(0)}(r)&=& 0~~, ~~~f_-^{(0)}(r)=0~,
\ee depending on which solution is normalizable.

One can determine the forms of the external field that admit
 exact solutions of the problem by comparing the forms of the
superpotential $W$ with those listed in Table~(4.1) of
\cite{Cooper}.

Similarly, from Turbiner's classification of the $sl(2)$ QES
systems \cite{Tur}, one can determine the forms of $W$, and hence
the forms of external fields admitting QES solutions based on
$sl(2)$ algebra.  The main ideas of the procedures are outlined
below.

\vskip 0.7cm

{\bf 3.} We shall concentrate only on solution of the upper
component $f_-$, which is assumed to have a normalizable zero
energy state.

Eq.(\ref{susy}) shows that $f_-$ satisfies the Schr\"odinger
equation $H_- f_-=\epsilon f_-$, with  \be H_-&=& A^-A^+
\nonumber\\ & =&-\frac{d^2}{dr^2} + V(r) ~,\label{SE}
\ee
with
\be
V(r)=W(r)^2 -W^\prime(r)~. \label{V}
\ee
We shall look for $V(r)$
such that the system is QES.  According to the theory of QES
models, one first makes an ``imaginary gauge transformation"
 on the function $f_-$
\be
f_-(r)= \phi(r) e^{-g(r)}~, \label{f-1} \ee where $g(r)$ is called
the gauge function.  The function $\phi(r)$ satisfies
\be
-\frac{d^2\phi(r)}{dr^2} + 2 g^\prime \frac{d\phi(r)}{dr} +
\left[V(r)+ g^{\prime\prime} - g^{\prime 2}\right]\phi
(r)=\epsilon\phi(r)~. \label{phi} \ee For physical systems which
we are interested in, the phase factor $\exp(-g(r))$ is
responsible for the asymptotic behaviors of the wave function so
as to ensure normalizability. The function $\phi(r)$ satisfies a
Schr\"odinger equation with a gauge transformed Hamiltonian
\be
H_G=-\frac{d^2}{dr^2} + 2W_0(r)\frac{d}{dr}  +\left[V(r)
+W_0^\prime - W_0^2\right]~, \label{HG} \ee where $W_0(r)=g^\prime
(r)$.  Now if $V(r)$ is such that the quantal system is QES, that
means the gauge transformed Hamiltonian $H_G$ can be written as a
quadratic combination of the generators $J^a$ of some Lie algebra
with a finite dimensional representation.  Within this finite
dimensional Hilbert space the Hamiltonian $H_G$ can be
diagonalized, and therefore a finite number of eigenstates are
solvable. For one-dimensional QES systems the most general Lie
algebra is $sl(2)$ .  Hence if Eq.(\ref{HG}) is QES then it can be
expressed as
\be
H_G=\sum C_{ab}J^a J^b + \sum C_a J^a + {\rm constant}~,
\label{H-g} \ee where $C_{ab},~C_a$ are constant coefficients, and
the $J^a$ are the generators of the Lie algebra $sl(2)$ given by
\be J^+ &=& z^2 \frac{d}{dz} - Nz~,\cr
J^0&=&z\frac{d}{dz}-\frac{N}{2}~,~~~~~~~~N=0,1,2\ldots\cr J^-&=&
\frac{d}{dz}~. \ee
 Here the variables $r$ and $z$ are related by
$z=h(r)$, where $h(\cdot)$ is some (explicit or implicit) function
. The value $j=N/2$ is called the weight of the differential
representation of $sl(2)$ algebra, and $N$ is the degree of the
eigenfunctions $\phi$, which are polynomials in a
$(N+1)$-dimensional Hilbert space with the basis $\langle
1,z,z^2,\ldots,z^N\rangle$:
\be
\phi=(z-z_1)(z-z_2)\cdots (z-z_N)~. \label{phi-2} \ee

The requirement in Eq.(\ref{H-g}) fixes $V(r)$ and $W_0(r)$, and
$H_G$ will have an algebraic sector with $N+1$ eigenvalues and
eigenfunctions.  For definiteness, we shall denote the potential
$V$ admitting $N+1$ QES states by $V_N$.   From Eqs.(\ref{f-1})
and (\ref{phi-2}), the function $f_-$ in this sector has the
general form
\be
f_-=(z-z_1)(z-z_2)\cdots (z-z_N)\exp\left(-\int^z W_0(r)
dr\right)~, \label{psi-1} \ee where $z_i$ ($i=1,2,\ldots,N$) are
$N$ parameters that can be determined by plugging Eq.(\ref{phi-2})
into Eq.(\ref{phi}).  The algebraic equations  so obtained are
called the Bethe ansatz equations corresponding to the QES problem
\cite{Ush,HoRoy1,HoRoy2} .  Now one can rewrite Eq.(\ref{psi-1})
as
\be
f_- =\exp\left(-\int^z W_N(r,\{z_i\}) dr\right)~, \label{f2} \ee
with
\be
W_N(r,\{z_i\}) = W_0(r) -  \sum_{i=1}^N
\frac{h^\prime(r)}{h(r)-z_i}~. \label{W} \ee There are $N+1$
possible functions $W_N (r,\{z_i\})$ for the $N+1$ sets of
eigenfunctions $\phi$. Inserting Eq.(\ref{f2}) into $H_-
f_-=\epsilon f_-$, one sees that $W_N$ satisfies the Ricatti
equation
\be
W_N^2 - W_N^\prime = V_N - \epsilon_N~, \label{Ricatti} \ee where
$\epsilon_N$ is the energy parameter corresponding to the
eigenfunction $f_-$ given in Eq.(\ref{psi-1}) for a particular set
of $N$ parameters $\{z_i\}$.

From Eqs.(\ref{SE}), (\ref{V}) and (\ref{Ricatti}) it is clear how
one should proceed to determine the external fields so that the
Dirac equation becomes QES based on $sl(2)$: one needs only to
determine the superpotentials $W(r)$ according to
Eq.(\ref{Ricatti}) from the QES potentials $V(r)$  classified in
\cite{Tur}. This is easily done by observing that the
superpotential $W_0$ corresponding to $N=0$ is related to the
gauge function $g(r)$ associated with a particular class of QES
potential $V(r)$ by $g^\prime (r)=W_0 (r)$.  This superpotential
gives the field configuration  that allows the weight zero
($j=N=0$) state, i.e. the ground state, to be known in that class.
The more interesting task is to obtain higher weight states (i.e.
$j>0$), which will include excited states.  For weight $j$
($N=2j$) states, this is achieved by forming the superpotential
$W_N(r,\{z_i\})$ according to Eq.(\ref{W}). Of the $N+1$ possible
sets of solutions of the Bethe ansatz equations, the set of roots
$\{z_1,z_2,\ldots,z_N\}$ to be used in Eq.(\ref{W}) is chosen to
be the set for which the energy parameter of the corresponding
state is the lowest.

\vskip 0.7cm

{\bf 4.} Let us illustrate the above procedure by an example. We
consider the motion of a neutral fermion of spin-1/2 with mass $m$
coupled non-minimally with an external electromagnetic field with
an anomalous magnetic moment $\mu$.  The relevant equation
describing such particle is the Dirac-Pauli equation \cite{D-P}.
This equation is useful in describing the celebrated
Aharonov-Casher effect \cite{AC-eff}, and is also of some interest
in quantum chromodynamics in connection with the problem of quark
confinement\cite{Dirac-osc}.

We shall consider the situation in which only electric field $\bf
E$ is present. In this case, the Dirac-Pauli equation $H\psi={\cal
E}\psi$ is described by the Hamiltonian
 \be
 H={\balpha}\cdot{\bf p}+i\mu{\bgamma}\cdot{\bf
E} +\beta m~\label{H}~,
 \ee
  with ${\bf p}=-i\nabla$ and $\beta=\gamma^0$.
 We choose the Dirac matrices in the
standard representation
\be
{\balpha} = \left( \begin{array}{cc} 0 & {\bsigma}\\ {\bsigma} & 0
\end{array}\right)~,~~~~~
\beta= \left( \begin{array}{cc} 1 & 0\\ 0 & -1
\end{array}\right)~,
\ee where $\bsigma$ are the Pauli matrices. We also define
$\psi=(\chi, \varphi)^t$, where $t$ denotes transpose, and both
$\chi$ and $\varphi$  are two-component spinors. Then the
Dirac--Pauli equation becomes
\be
{\bsigma}\cdot({\bf p}-i\mu {\bf E})\chi &=&({\cal
E}+m)\varphi~,\nonumber\\
 {\bsigma}\cdot({\bf p}+i\mu {\bf
E})\varphi &=&({\cal E}-m)\chi~.\label{H1} \ee

We now consider central electric field ${\bf E}=E_r {\hat{\bf
r}}$. In this case, one can choose a complete set of observables
to be $(H,{\bf J}^2,J_z,{\bf S}^2=3/4,K)$. Here $\bf J$ is the
total angular momentum ${\bf J=L+S}$, where $\bf L$ is the orbital
angular momentum, and ${\bf S}=\frac{1}{2}{\bf\Sigma}$ is the spin
operator. The operator $K$ is defined as
$K=\gamma^0({\bf\Sigma}\cdot{\bf L}+1)$, which commutes with both
$H$ and {\bf J}. Explicitly, we have
\be
K&=&{\rm diag }\left({\hat k},-{\hat k}\right)~,\nonumber\\ {\hat
k}&=&\bsigma\cdot {\bf L} +1~. \ee
 The common eigenstates can be
written as
\be
\psi=\frac{1}{r} \left( \begin{array}{c} f_-(r) {\cal
Y}^k_{jm_j}\\ if_+(r){\cal Y}^{-k}_{jm_j}
\end{array}\right)~,
\ee here ${\cal Y}^k_{jm_j}(\theta,\phi)$ are the spin harmonics
satisfying \be {\bf J}^2 {\cal Y}^k_{jm_j} &=& j(j+1){\cal
Y}^k_{jm_j}~,~~j=\frac{1}{2},\frac{3}{2},\ldots~~,
\\ J_z {\cal Y}^k_{jm_j} &=& m_j{\cal
Y}^k_{jm_j}~,~~~~~~~~|m_j|\leq j~~,
\\ {\hat k}{\cal Y}^k_{jm_j}&=& -k{\cal Y}^k_{jm_j}~,~~~~~~~~~
k=\pm(j+\frac{1}{2})~,  \ee and
\be
({\bsigma}\cdot {\hat {\bf r}}){\cal Y}^k_{jm_j}= -{\cal
Y}^{-k}_{jm_j}~, \ee  where $\hat{\bf r}$ is the unit radial
vector. Eq.(\ref{H1}) then reduces to
\be
\left(\frac{d}{dr} + \frac{k}{r} + \mu E_r\right)f_- &=&
\left({\cal E} + m\right)f_+~,\label{f-}\\ \left(-\frac{d}{dr} +
\frac{k}{r} + \mu E_r\right)f_+ &=& \left({\cal E} - m\right)f_-~.
\label{f+} \ee This shows that $f_-$ and $f_+$ forms a
one-dimensional SUSY pairs with the superpotential $W$ given by
\be
W=\frac{k}{r} + \mu E_r~, \label{W1} \ee and the energy parameter
$\epsilon={\cal E}^2 -m^2$.

We can now classify the forms of the electric field $E_r(r)$ which
allow exact and quasi-exact solutions. To be specific, we consider
the situation where $k<0$ and $\int dr \mu E_r >0$, so that
$f_-^{(0)}$ is normalizable, and $f_+^{(0)}=0$. The other
situation can be discussed similarly. In this case, Eq.(\ref{W1})
becomes
\be
W=-\frac{|k|}{r} + \mu E_r~. \label{W2}
 \ee

We determine the forms of $E_r$ that give exact/quasi-exact energy
$\cal E$ and the corresponding function $f_-$.  The corresponding
function $f_+$ is obtained using Eq.(\ref{f-}).

\vskip 0.7cm

{\bf 5.}  Comparing the forms of the superpotential $W$ in
Eq.(\ref{W2}) with Table~(4.1) in \cite{Cooper}, one concludes
that there are three forms of $E_r$ giving exact solutions of the
problem :

~~~~~~~~~~i)   oscillator-like :  $\mu E_r(r)\propto r $~;

~~~~~~~~~~ii)  Coulomb potential-like :   $\mu E_r(r)\propto {\rm
constant} $~;

~~~~~~~~~~iii) zero field-like :  $\mu E_r(r)\propto 1/r $~.

Case (i) and (ii) had been considered in \cite{SV} and \cite{Lin},
and case (iii) in \cite{SV}.

We mention here that the case with oscillator-like field, i.e.
case (i), is none other than the spherical Dirac oscillator
\cite{Dirac-osc}.

\vskip 0.7cm

{\bf 6.} The form of the superpotential $W$ in Eq.(\ref{W2}) fits
into three classes, namely, Classes VII, VIII and IX of
$sl(2)$-based QES systems in \cite{Tur}.  Below we shall
illustrate our construction of QES electric fields in Class VII
QES systems.

The general potential in Class VII has the form
\be
V_N(r)=a^2r^6+2abr^4+\left[b^2-a\left(4N+2\gamma
+3\right)\right]r^2
+\gamma\left(\gamma-1\right)r^{-2}-b\left(2\gamma+1\right)~,
\label{cVII} \ee where $a,b$ and $\gamma$ are constants. The gauge
function is
\be
g(r)=\frac{a}{4}r^4 + \frac{b}{2}r^2 -\gamma\ln {r}~. \label{g4}
\ee We must have $a,\gamma >0$ to ensure normalizability of the
wave function.  Eqs.(\ref{g4}) and (\ref{W2}), together with the
relation $W_0(r)=g^\prime(r)$, give us the electric field
$E_r^{(0)}$:
\be
\mu E_r^{(0)} (r)= ar^3 + b r~. \ee The Dirac-Pauli equation with
this field configuration admits a QES ground state with energy
${\cal E}^2=m^2$ ($\epsilon=0$) and ground state function
$f_-\propto \exp(-g_0(r))$. Also, here we have $\gamma=|k|$.

To determine electric field configurations admitting QES
potentials $V_N$ with higher weight, we need to obtain the Bethe
ansatz equations for $\phi$. Letting $z=h(r)=r^2$, Eq.(\ref{phi})
becomes
\begin{eqnarray}
\left[-4 z \frac{d^2}{dz^2} +\left(4az^2 +4bz -2\left(2\gamma
+1\right) \right)\frac{d}{dz} -\left(4aNz + \epsilon
\right)\right]\phi(z) =0~. \label{phi-VII}
\end{eqnarray}
For $N=0$, the value of the $\epsilon$ is $\epsilon=0$. For higher
$N>0$ and $\phi(r)=\prod_{i=1}^N (z-z_i)$, the electric field
$E_r^{(N)}(r)$ is obtained from Eq.(\ref{W}):
\begin{eqnarray}
\mu E_r^{(N)}(r) = \mu E_r^{(0)}(r) - \sum_{i=1}^N
\frac{h^\prime(r)}{h(r)-z_i}~. \label{EN}
\end{eqnarray}
For the present case, the roots $z_i$'s are found from the Bethe
ansatz equations
\begin{eqnarray}
2az_i^2 +2bz_i -\left(2\gamma+1\right) - \sum_{l\neq
i}\frac{z_i}{z_i-z_l} =0~, \quad\quad i=1,\ldots,N~,
\label{BA-VII}
\end{eqnarray}
and $\epsilon$ in terms of the roots $z_i$'s is
\begin{eqnarray}
\epsilon=2\left(2\gamma+1\right)\sum_{i=1}^N \frac{1}{z_i}~.
\label{E-VII}
\end{eqnarray}

For $N=1$ the roots $z_1$ are
\begin{eqnarray}
z_1^\pm=\frac{-b\pm\sqrt{b^2+2a(2\gamma+1)}}{2a}~, \label{z1}
\end{eqnarray}
and the values of $\epsilon$ are
\begin{eqnarray}
\epsilon^\pm=2\left(b\pm\sqrt{b^2+2a(2\gamma+1)}\right)~.
\end{eqnarray}
 For $a>0$, the root $z_1^-=-|z_1^-|<0$ gives the
ground state. With this root, one gets the superpotential
\begin{eqnarray}
W_1(r)=ar^3 +br -\frac{2r}{r^2+|z_1^-|} -\frac{\gamma}{r}~.
\end{eqnarray}
From Eq.(\ref{EN}), the corresponding electric field is
\be
\mu E_r^{(1)} (r) &=& ar^3 +br - \frac{2r}{r^2+ |z_1^-|}~. \ee The
QES potential appropriate for the problem is
\begin{eqnarray}
V(x) &=& W_1^2-W_1^\prime~,\cr &=&V_1 -\epsilon~.
\end{eqnarray}
The one-dimensional SUSY sets the energy parameter of ground state
at $\epsilon=0$.  Hence, the ground state and the excited state
have energy parameter $\epsilon=0$ and $\epsilon=\epsilon^+
-\epsilon^-=4\sqrt{b^2+2a(2\gamma+1)}$, and wave function
\be
f_-\propto e^{-g_0(r)}\left(r^2-z_1^-\right) \ee and
\be
f_-\propto e^{-g_0(r)}\left(r^2-z_1^+\right)~, \ee respectively.

QES potentials and electric fields for higher degree $N$ can be
constructed in the same manner.

A more extensive discussion of the (quasi)-exact solvability of
the Dirac equation in different background potentials can be found
in \cite{BN}.

\bigskip
{\small This work was supported in part by the National Science
Council of the Republic of China through Grant No. NSC
94-2112-M-032-007 and NSC 96-2112-M-032-007-MY3. }

\end{document}